\documentclass[conference,twocolumn, 10pt, letterpaper]{IEEEtran}

\makeatletter
\def\ps@headings{%
\def\@oddhead{\mbox{}\scriptsize\rightmark \hfil \thepage}%
\def\@evenhead{\scriptsize\thepage \hfil \leftmark\mbox{}}%
\def\@oddfoot{}%
\def\@evenfoot{}}
\makeatother 

\usepackage{epsfig, latexsym, amssymb, amsmath, url}
\usepackage{graphics}
\usepackage{algorithm}
\usepackage{algorithmic}
\usepackage{subfigure}

\newtheorem{prop}{Proposition}[section]

\newtheorem{cor}{Corollary}

\newtheorem{lm}{Lemma}

\newtheorem{thm}{Theorem}

\newcommand{\bthm}{\begin{thm}}
\newcommand{\ethm}{\end{thm}}

\newcommand{\bcor}{\begin{cor}}
\newcommand{\ecor}{\end{cor}}
\newcommand{\bprop}{\begin{prop}}
\newcommand{\eprop}{\end{prop}}
\newcommand{\blm}{\begin{lm}}
\newcommand{\elm}{\end{lm}}
\newcommand{\beq}{\begin{equation}}
\newcommand{\eeq}{\end{equation}}
\newcommand{\ber}{\begin{eqnarray}}
\newcommand{\eer}{\end{eqnarray}}

\newenvironment{proof1}{\begin{trivlist}\item[]{\bf Proof:\hspace{2mm}}}{\hfill$\blackbox$\end{trivlist}}

\newcommand{\blackbox}{\vrule height7pt width5pt depth1pt}

\newcommand{\bit}{\begin{itemize}}
\newcommand{\eit}{\end{itemize}}
\newcommand{\ben}{\begin{enumerate}}
\newcommand{\een}{\end{enumerate}}
\newcommand{\bdesc}{\begin{description}}
\newcommand{\edesc}{\end{description}}
\newcommand{\beqarrn}{\begin{eqnarray*}}
\newcommand{\eeqarrn}{\end{eqnarray*}}
\newenvironment{proofof}[1]{\begin{trivlist}\item[]{\bf Proof of #1:\hspace{2mm}
}}{\hfill\blackbox\end{trivlist}}
\newcommand{\bproofof}{\begin{proofof}}
\newcommand{\eproofof}{\end{proofof}}
\newenvironment{rem}{\begin{trivlist}\item[]{\bf
Remark:}\hspace{4mm}}{\end{trivlist}}
\newcommand{\brem}{\begin{rem}}
\newcommand{\erem}{\end{rem}}
\newenvironment{rems}{\begin{trivlist}\item[]{\bf
Remarks}\begin{itemize}}{\end{itemize}\end{trivlist}}
\newcommand{\brems}{\begin{rems}}
\newcommand{\erems}{\end{rems}}
\newtheorem{fact}{Fact}
\newcommand{\bfact}{\begin{fact}}
\newcommand{\efact}{\end{fact}}
\newtheorem{examp}{Example}
\newcommand{\bexamp}{\begin{examp}\rm}
\newcommand{\eexamp}{\end{examp}}
\newtheorem{defn}{Definition}
\newcommand{\bdefn}{\begin{defn}\rm}
\newcommand{\edefn}{\end{defn}}

\newtheorem{prob}{Problem}
\newcommand{\bprob}{\begin{prob}}
\newcommand{\eprob}{\end{prob}}

\newcommand{\bvtm}{\begin{verbatim}}
\newcommand{\bfig}{\begin{figure}}
\newcommand{\efig}{\end{figure}}
\newcommand{\bcen}{\begin{center}}
\newcommand{\ecen}{\end{center}}

\long\def\comment#1{}

\def \n2{{N_0 \over 2}}

\def \h5{\hspace{0.5in}}

\begin{document}

\title{\ Diffusion of Real-Time Information in Social-Physical Networks}
\author{\IEEEauthorblockN{Dajun Qian\IEEEauthorrefmark{1},
Osman Ya\u{g}an\IEEEauthorrefmark{2},  Lei Yang\IEEEauthorrefmark{1}
and  Junshan
Zhang\IEEEauthorrefmark{1}   }

\IEEEauthorblockA{\IEEEauthorrefmark{1} School of ECEE,
Arizona State University, Tempe, AZ, USA\\
} \IEEEauthorblockA{\IEEEauthorrefmark{2} Cybersecurity Laboratory (CyLab), Carnegie Mellon University, Pittsburgh, PA, USA \\}}

\maketitle
 \pagestyle{empty}
  \thispagestyle{empty}

\begin{abstract}
We study the diffusion behavior of real-time information. Typically,  real-time information  is
valuable only for a limited time duration,  and hence needs to be delivered  before its ``deadline.''
Therefore, real-time information is much easier to spread among a group of
people with frequent interactions than between isolated individuals. With this insight, we consider
a social network  which consists of many   cliques and   information can spread quickly within
a clique.  Furthermore, information can also be shared through online social networks, such as
Facebook, twitter, Youtube, etc.

We characterize the diffusion of real-time information  by studying the phase transition
behaviors. Capitalizing on the theory of inhomogeneous random networks, we show that the social network has a critical threshold above which information epidemics are very likely to happen. We also theoretically quantify
the fractional size of individuals that finally receive the message. The  numerical results indicate that
real-time information could be much easier to propagate in a social network when large size  cliques exist.
\end{abstract}

\section{Introduction}
\subsection{Motivation and Background}
In today's modern society, people are becoming increasingly tied together over social networks.
Thanks to \emph{online social networks}, such as Facebook and Twitter, people can share
messages quickly with their   friends \cite{mislove2007measurement}. Meanwhile,
 a \emph{physical information
network} \cite{leskovec2007dynamics,isella2010s,zhao2011social}   based on
traditional face-to-face interactions still remains an important medium for  message spreading. Very recent work
\cite{osmanmessage} has shown that different social networks are usually \emph{coupled} together, and the
conjoining could greatly facilitate  information diffusion. As a result,  today's hot spot news or fashion
behaviors are more likely to generate pronounced influence over the population than ever before.

The main thrust of this study is dedicated to   understanding  the diffusion behavior of real-time
information. Typically, the real-time information is   valuable only for a limited time duration
\cite{yang2010modeling}, and hence needs to be delivered  before its ``deadline.''
 For example, once a limited-time coupon is released from Groupon or
Dealsea.com, people can share this news  either by talking to  friends or
 posting it on Facebook. However, people would not have much interest on this deal after it is not longer available.

Clearly, due to the timeliness requirement, the  potential influenced
scale of real-time information in a social network depends on the speed of message propagation. The faster the
message passes from one to another, the more people can learn this news before it expires. With
this insight, in order to   characterize its diffusion behavior, a key step is to quantify how
fast the message can spread along different social connections.

In this study, we assume that   information could spread amongst people through both face-to-face
contacts and online communications. In an online social network, the message can spread quickly
over long distance, and hence it is reasonable to  treat   online connections   as the same type of
links regardless of real-world distances.

On the other hand, face-to-face communications are largely constrained by  distance between
individuals.   Recent works in \cite{zhao2011social,stehlé2010dynamical} have explored the
structure of physical information network by tracking in-person interactions over the population. Their
findings indicate that such interactions would give rise to a social graph made of a large number
of small isolated \emph{cliques}. Each clique stands for a group of  people who are close to each
other. The message can spread quickly within a clique via   frequent interactions, but takes longer
time  to spread across cliques   separated by   longer distances. Clearly,   constrained by its
deadline, the real-time information could be less likely to propagate across cliques via
face-to-face contacts. Needless to say, in order to characterize the diffusion behavior of real-time
information, we need to consider the impact of such clique structure, which has not been studied in
  previous works on general information diffusions
\cite{osmanmessage,newman2001random, newman2002spread}.


\subsection{Summary of Main Contributions}
We explore the diffusion  of real-time information in an overlaying social-physical network
 where the information could spread amongst people through both face-to-face contacts (physical information network) and online communications (online social network).
 Based on empirical observations in  \cite{zhao2011social,stehlé2010dynamical},  we assume that the physical information
network consists of many isolated cliques where each clique represents a group of people with
frequent face-to-face interactions, e.g., family in a house or colleagues in an office. Clearly,
the face-to-face contacts are less likely to happen  across cliques.

We characterize the information diffusion process  by  studying the  \emph{phase transition
behaviors} (see Section~\ref{sec:information_cascade}
for details). Specifically, we show that the social network has a \emph{critical threshold} above which {\em
information epidemics} are very likely to happen, i.e., the information can reach a non-trivial
fraction of individuals. We also quantify the fraction of individuals that finally receive the
message by computing the size of \emph{giant component} (see Section~\ref{sec:information_cascade}
for details).  The numerical results in Section~\ref{sec:numerical} indicate that
real-time information could be much easier to propagate in a social network when large size  cliques exist.
As illustrated in Figure~\ref{fig:nodesize}, when the average clique size increases from $1$ to $2$,
the fraction of individuals that receive the message can grow sharply from $14\% $ to $80\%$.

Note that our work here has significant differences from the previous studies on information
propagation. In \cite{newman2001random,newman2002spread}, it is assumed that message could
propagate at the same speed along different social relationships.  Clearly, such assumption would
be inappropriate for the diffusion of real-time information that depends on propagation speeds.
Very recent work in \cite{osmanmessage} considered online connections and face-to-face connections
for general information diffusion, but did not study the impact of clique structure on information
diffusion. To the best of our knowledge, this paper is the first work on the diffusion of real-time
information with consideration on the clique structure in  social networks. We believe that our
work  will   offer initial steps towards understanding  the  diffusion behaviors of real-time
information.

\section{System model} \label{sec:model}
We consider an overlying social-physical network   $\mathbb{H}$ that consists of a physical information network
$\mathbb{W}$ and an online social network $\mathbb{F}$. The nodes in  the graph $\mathbb{W}$
represent the human beings in the real world. Based on empirical studies in
\cite{zhao2011social,stehlé2010dynamical}, we assume that the graph $\mathbb{W}$ consists of
 many isolated cliques
where each clique represents a group of closely located people, e.g., family in a house or
colleagues in an office.
 Meanwhile, each node in $\mathbb{W}$ can independently
participate the online social network $\mathbb{F}$ with  probability $\alpha$, and  the   nodes in
$\mathbb{F}$ stand for their online memberships. Throughout this paper, we also refer to the nodes
in $\mathbb{W}$ and $\mathbb{F}$ as ``individuals'' and ``online users,'' respectively.
Furthermore, the links connecting the nodes in $\mathbb{W}$ stand for traditional face-to-face
connections, while the links in $\mathbb{F}$ represent online connections.
\begin{figure}
\begin{center}
\includegraphics[totalheight=0.19\textheight,
width=.47\textwidth]{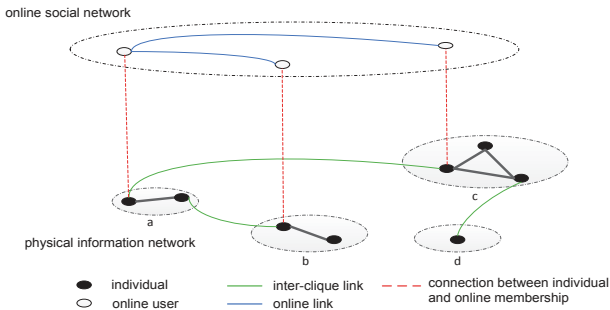} \caption{\sl System model $\mathbb{H}$ } \label{fig:model}
\end{center} \vspace{-0.8cm}
\end{figure}

\subsection{Topology Structure in System Model}
In what follows, we specify the topology structure in the  system model $\mathbb{H}$  in
Fig.~\ref{fig:model}.

\textbf{Cliques in the physical information network}. The physical information network has  $N$ nodes  and
the nodes set is denoted by $ \mathcal{N}= \{1,2,...,N\}$. These nodes are gathered into many
cliques with different sizes.
 We expect the clique size follows the distribution $\{ \mu _n^w,
n=1,2,...,D \}$, $n=1,2,...D$, where $D$ is the largest possible size. Therefore, an arbitrary
clique could contain $n$ nodes  with probability  $\mu _n^w$.  We generate these cliques as
follows: at step $t=1$, we randomly choose $n$ nodes from the    collection $\mathcal{N}$ and
create a clique with the selected $n$ nodes, where $n$ is a random number following the
distribution  $\{ \mu _n^w, n=1,2,...,D \}$. We also denote the collection of the remaining nodes
in $ \mathcal{N}$ by $\mathcal{N}_1$. At each step $t$, we repeat the above procedure to create a
new clique from the collection  $\mathcal{N}_{t-1}$\footnote{Note that the  last generated clique
may not follow the expected size distribution, since  there would be only too few nodes left to
choose. However, such impact on clique size distribution will be negligible if the number of
cliques is large enough.}, and  assume that we can finally generate $N_c$ cliques in $\mathbb{W}$
\footnote{Throughout this paper, we use ``clique in $\mathbb{W}$'' and ``clique in $\mathbb{H}$''
interchangeably, in the sense that the network  $\mathbb{W}$ is also a part of  system model
$\mathbb{H}$.}.
 Generally speaking, the existence of large
size cliques indicates that many individuals are close to each other. In other words, the clique
size distribution $\{ \mu _n^w \}$ offers an abstract characterization of  personal distances in $\mathbb{W}$
from a macroscopic perspective.

\textbf{Type-$0$ (intra-clique) links in $\mathbb{W}$}.  Since the nodes within the same cliques
could interact to each other frequently, we assume these nodes are fully  connected by \emph{type-$0$ links}.

\textbf{Type-$1$ (inter-clique) links  in $\mathbb{W}$}.  We assume that a face-to-face interaction is still possible
between  cliques, e.g., a person may talk to a remote friend by walking across a long distance.
 Suppose  each node can randomly connect to $k^w$ nodes from other
cliques through \emph{type-$1$ links} where $k^w$ is a random variable drawn independently from the distribution $\{ {p_k^w, k=0,1,...} \}$.

\textbf{Online users and type-$2$ (online) links}. The nodes in the online network $\mathbb{F}$
represent the online users.  As in \cite{osmanmessage}, we assume each online user randomly
connects to $k^f$ online neighbors in $\mathbb{F}$, where $k^f$ is a random variable whose distribution is drawn
independently from $\{p_k^f, k=0,1,...\}$. We denote such online connection as \emph{type-$2$ link}.
Furthermore, we draw a virtual \emph{type-$3$  link} from an online user in  $\mathbb{F}$ to the
actual person it corresponds to in the physical information network $\mathbb{W}$;  this indicates that
the two nodes actually correspond  to the same individual.

%

\textbf{Online users associated with a clique}. To avoid confusions, we say \emph{``size-$n$ clique
with $m$ online members''} when referring to the case that a clique contains $n$ individuals and
only $m$ of them participate in the online social network  $\mathbb{F}$. With this insight, we can
also differentiate among the collection of  size-$n$ cliques according to their affiliated online
users. Specifically, for the collection of size-$n$ cliques with $m$ online members, $m \le n \le
D$, we assume their  fraction size  in the whole collection of cliques is $\mu _{nm}$. It is easy to
see that
\begin{equation}
{\mu _{nm}} = \mu _n^w\left( {\begin{array}{*{20}{c}}
n\\
m
\end{array}} \right){\alpha ^m}{\left( {1 - \alpha } \right)^{n - m}} ~~ \textrm{and} ~~ {\mu_n^w} = \sum\limits_{m= 1}^n {\mu _{nm}^{}}.
\end{equation}
Furthermore, for the collection of cliques with $m$ online users,  their fraction size  can be given by
\begin{equation} \label{eq:u_m} {  \mu_m^f} = \sum\limits_{n=m}^D {\mu _{nm}^{}}.
\end{equation}

\subsection{Information Transmissibility}
The message can  propagate at different speeds along different types of social connections in  $\mathbb{H}$.
 Due to timeliness requirement, the real-time information is
easier to pass over a link that offers fast propagation speed. Therefore, we assign each link  with
a \emph{transmissibility} as in \cite{osmanmessage,newman2002spread}, i.e., the probability that
the message can successfully pass through.

From practical scenarios,  we set the transmissibility along type-$0$ link as $T_c=1$  since the
message spreads quickly within a clique. We also define  the transmissibilities along type-$1$ and
type-$2$ links as $T_w$ and $T_f$, respectively. Throughout this paper, we say a link is
\emph{occupied} if the message can successfully pass through that link. Hence, in $\mathbb{H}$ each
type-$1$ link is occupied independently with probability $T_w$, whereas each type-$2$ link is
occupied independently with probability $T_f$.

\subsection{Information Cascade}
\label{sec:information_cascade}

We  give a brief description of  the information diffusion process in the following. Suppose that
the message starts to spread from an arbitrary  node $i$ in a clique of $\mathbb{W}$. Then, the
other nodes  in this clique will quickly receive that message through type-$0$ links. The message
can also propagate to other cliques through occupied type-$1$ and type-$2$ links. This in turn may
trigger further message propagation and may eventually lead to an information epidemic; i.e., a
non-zero fraction of individuals may receive the information in the limit $N \to \infty$.

Clearly, an arbitrary individual can spread the information to nodes that are reachable from itself
via the occupied edges of $\mathbb{H}$. Hence, the size of an information outbreak (i.e., the
number of individuals that are informed) is closely related to the size of the connected components
of $\mathbb{{H}}$, which contains only the \emph{occupied} type-$1$ and type-$2$ links
\cite{osmanmessage,newman2002spread,newman2001random} of $\mathbb{H}$. Thus, the information
diffusion process considered here is equivalent to a heterogeneous bond-percolation process over
the network $\mathbb{H}$; the corresponding bond percolation is heterogeneous since the occupation
probabilities are different for type-$1$ and type-$2$ links. In this paper, we will exploit this
relation and find the condition and the size of information epidemics by studying the phase
transition properties of $\mathbb{{H}}$. A key observation is that the system $\mathbb{{H}}$
exhibits a  \emph{phase transition} behavior  at a \emph{critical threshold}. Specifically, a
\emph{giant connected component} $G_H$  that covers a non-trivial fraction of nodes in
$\mathbb{{H}}$ is likely to appear  above the critical threshold meaning that \emph{information
epidemics} are possible. Below that critical threshold, all components are small indicating that
the fraction of influenced individuals tends to zero in the large system size limit.

It is easy to see that the influenced
individuals and cliques correspond to the nodes and cliques in $\mathbb{W}$ that are contained
inside   $G_H$. Hence, we introduce two parameters to evaluate the scale of information diffusion:
\begin{itemize}
  \item $S_c$:  The fractional size of influenced cliques in $\mathbb{W}$.
  Namely, $S_c$ corresponds to the ratio of the number of cliques in $G_H$ to the total number of cliques in $\mathbb{W}$.
  \item $S_n$: The fractional size of influenced individuals in  $\mathbb{W}$. Namely,
  $S_n$ corresponds to the ratio of the number of nodes that belong to the cliques in $G_H$ to
the total number of nodes in $\mathbb{W}$.
\end{itemize}
With this insight, we can explore the information diffusion process by
 characterizing the phase transition behavior of the giant component $G_H$.

\section{Equivalent graph: a  clique level   approach} \label{sec:equal_graph}
In this study, we are particularly interested in  the following two questions:
\begin{figure}[!t]
\begin{center}
\includegraphics[totalheight=0.15\textheight,
width=.3\textwidth]{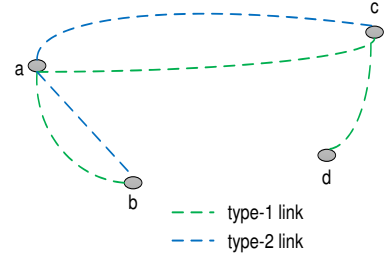}\caption{Equivalent graph $\mathbb{E}$. Nodes \{a,b,c,d\} in this
graph
 corresponds to the cliques \{a,b,c,d\} of the system $\mathbb{H}$ in Fig.~\ref{fig:model}.
We assign type-$1$ and type-$2$ links in  $\mathbb{E}$ according to the same type of links
connecting cliques in Fig.~\ref{fig:model}.} \label{fig:equal}
\end{center} \vspace{-0.5cm}
\end{figure}
\begin{itemize}
  \item   What is the critical threshold of the system $\mathbb{H}$? In other words,
under what condition, the information reaches a non-trival fraction of the network rather than dying
out quickly?
  \item What is the {\em expected} size of an information epidemic? In other words,
  to what fraction of nodes and cliques does the information reach? Or, equivalently,
  what are the sizes $S_c$ and $S_n$?
\end{itemize}
 These two questions can be  answered by quantifying  the phase transition behaviors
of the random graph $\mathbb{H}$. Due to the clique structure in our system model,
 the techniques employed in  existing works \cite{osmanmessage,newman2001random,newman2002spread}
 cannot be directly applied here.
To tackle this challenge, we develop an equivalent random graph $\mathbb{E}$ that exhibits the same
phase transition behavior as the original model $\mathbb{H}$. Then, we characterize the
phase transition behaviors in the graph $\mathbb{E}$  by capitalizing on the recent results in
\emph{inhomogeneous random graph} \cite{Soderberg1,Soderberg2}.

We first  construct  an equivalent graph $\mathbb{E}$ based on the  topology structure of $\mathbb{H}$. Since
the nodes within the same clique can immediately share the message,
 we treat each clique including affiliated online users as
a single  virtual node in $\mathbb{E}$. Furthermore, we assign type-$1$ and type-$2$ links between
two virtual nodes  according to the original connections in $\mathbb{H}$. To get a more concrete
sense, we depict  the equivalent graph in Fig.~\ref{fig:equal} that corresponds to the original
model in Fig.~\ref{fig:model}. It is easy to see that the (type-$1$ and type-$2$)  link degree  of
a virtual node equals the total number of (type-$1$ and type-$2$)  links that are incident on the
nodes within the corresponding clique.  The equivalent graph $\mathbb{E}$ is expected to
exhibit the same phase transition behavior as the original model $\mathbb{H}$ since both
graphs have the
similar connectivity structure. In particular,  the fractional size of the giant component $G_E$
in the equivalent graph $\mathbb{E}$ (the ratio of the number of nodes in $G_E$ to the number of nodes
in  $\mathbb{E}$) is equal to the aforementioned quantity $S_c$. Thus, with a slight abuse of notation, we
use $S_c$ to denote the fractional size of $G_E$.

 The degree of an arbitrary node  in $\mathbb{E}$ can be
represented by a two-dimensional vector $\boldsymbol{d} =[d^w \:\: d^f]$
 where $d^w$ and $d^f$ correspond to the numbers of type-$1$ and   type-$2$ links incident on that node, respectively.   For a node in $\mathbb{E}$ that
corresponds to a size-$n$ clique  in $\mathbb{W}$, we use  $K_n^w$ to denote its  type-$1$ link degree,
 where $K_n^w$ is a random variable following the distribution $\{ P_{nk}^w, k=0,1,2,...\}$.
 Similarly, for a node in $\mathbb{E}$ that corresponds to a clique with $m$ online users,   we use  $K_m^f$ to denote its type-$2$ link degree where  $K_m^f$ follows the distribution $\{ P_{mk}^f, k=0,1,2,...\}$.
It is clear to see that an arbitrary node in $\mathbb{E}$ has link degree $[i \:\: j]$ with probability
\begin{equation}
p(i,j)  = \sum\limits_{n = 1}^D {\sum\limits_{m = 0}^n {{\mu _{nm}}P_{ni}^wP_{mj}^f} } ~~~   i,j
\in N. \end{equation}
Let  ${\rm{E}}[{d_w}]$ and ${\rm{E}}[{d_f}]$ be the mean numbers of type-$1$ and type-$2$ links   for a node in $\mathbb{E}$, i.e., ${\rm{E}}[{d_w}] = \sum\nolimits_{i = 0}^\infty  {\sum\nolimits_{j = 0}^\infty  {p(i,j)i} }$ and
${\rm{E}}[{d_f}] = \sum\nolimits_{i = 0}^\infty  {\sum\nolimits_{j = 0}^\infty  {p(i,j)j} }$. We also define ${\rm{E}}[{d_w}{d_f}] = \sum\nolimits_{i = 0}^\infty  {\sum\nolimits_{j = 0}^\infty  {p(i,j)ij} }$.
Furthermore, let ${\rm{E}}[{({d_w})^2}]$ and ${\rm{E}}[{({d_f})^2}]$  denote the second moments of the number of type-$1$ and type-$2$ links for  a node in $\mathbb{E}$, respectively; i.e., ${\rm{E}}[{({d_w})^2}] = \sum\nolimits_{i = 0}^\infty  {\sum\nolimits_{j = 0}^\infty  {p(i,j){i^2}} }$ and ${\rm{E}}[{({d_f})^2}] = \sum\nolimits_{i = 0}^\infty  {\sum\nolimits_{j = 0}^\infty  {p(i,j){j^2}} }$.

\section{Analytical solutions}
\label{sec:analysis} In this section, we analyze   information diffusion process by
characterizing the phase transition behaviors in the equivalent random graph  $\mathbb{E}$. We
present our analytical results in the following two steps. We first quantify the  conditions for
the emergence of  a giant component as well as the fractional sizes  $S_c$ and  $S_n$ for the
special case $T_w=1$ and $T_f=1$. We next show that these results can be easily extended to
a more general case with $0 \le T_w \le 1$ and $0 \le T_f \le 1$.

\subsection{Special Case: $T_w=T_f=1$}  \label{sec:case1}
We characterize the phase transition behavior of the giant component in $\mathbb{E}$  by
capitalizing the theory of inhomogeneous random graphs  \cite{Soderberg1,Soderberg2,Soderberg4}.
Specifically,  we define
 ${a_{11}} = {\rm{E[(}}{d_w}{{\rm{)}}^2}{\rm{]/E[}}{d_w}{\rm{]}} - 1$,
 ${a_{12}} = {\rm{E[}}{d_w}{{\rm{d}}_f}{\rm{]/E[}}{d_w}{\rm{]}}$,
 ${a_{21}} = {\rm{E[}}{d_w}{{\rm{d}}_f}{\rm{]/E[}}{d_f}{\rm{]}}$ and
 ${a_{22}} = {\rm{E[(}}{d_f}{{\rm{)}}^2}{\rm{]/E[}}{d_f}{\rm{]}} - 1$. Along the same line in \cite{osmanmessage,Soderberg1,Soderberg4}, we   have the following result.
 \blm \label{lm:threshold}
Let
\begin{equation} \label{eq:threshold}
\sigma = \frac{1}{2}\left( {{a_{11}} + {a_{22}} + \sqrt {{{({a_{11}} - {a_{22}})}^2} + 4{a_{12}}{a_{21}}} } \right)
\end{equation}
 if $\sigma >1$,  with high probability (whp) there exists a giant component in $\mathbb{E}$, i.e., a
non-trival fraction of nodes in $\mathbb{E}$ are connected;
otherwise, a giant component does no exist in $\mathbb{E}$ whp.
 \elm

The proof of Lemma \ref{lm:threshold} is relegated to Appendix A. As we discussed in Section \ref{sec:information_cascade}, the
existence of a giant component in  $\mathbb{E}$ indicates  that the information can reach a non-trival
fraction of  cliques in $\mathbb{H}$ rather than dying out quickly.

Next, let $h_1$ and $h_2$ in $(0,1]$ be given by the smallest solution to the following recursive equations:
\begin{equation}    \label{eq:h1}
 {h_1} = \frac{1}{{ {{\rm{E}}[{d_w}]} }}\sum\limits_{n = 1}^D {\sum\limits_{m = 0}^n {{\mu _{nm}}{\rm{E}}[K_n^wh_1^{K_n^w - 1}]{\rm{E}}[h_2^{K_m^f}]} },
\end{equation}
 \begin{equation} \label{eq:h2}
 {h_2} = \frac{1}{{{\rm{E}}[{d_f}] }}\sum\limits_{n = 1}^D {\sum\limits_{m = 0}^n {{\mu _{nm}}{\rm{E}}[h_1^{K_n^w}]{\rm{E}}[K_m^fh_2^{K_m^f - 1}]} }.
\end{equation}

We  have the following results on the size and probability of an information epidemic.
 \blm \label{lm:size}
The fractional size of the giant component  in  $\mathbb{E}$ (equivalently, the fractional size of influenced cliques in $\mathbb{W}$) is given by
\begin{equation}\label{eq:gc}
{S_c} = \sum\limits_{n = 1}^D {\sum\limits_{m = 0}^n {{\mu_{nm}}\left( {1 -
{\rm{E}}[h_1^{K_n^w}]{\rm{E}}[h_2^{K_m^f}]} \right)} }.
\end{equation}
The fractional size of influenced nodes in $\mathbb{W}$ is given by
\begin{equation}\label{eq:node}
{S_n} = \frac{1}{C}\sum\limits_{n = 1}^D {\sum\limits_{m = 0}^n {n{\mu _{nm}}\left( {1 - {\rm{E}}[h_1^{K_n^w}]{\rm{E}}[h_2^{K_m^f}]} \right)} },
\end{equation}
with the normalization term $C = \sum\limits_{n = 1}^D {n{\mu _n}}$.
\elm
The proof of Lemma \ref{lm:size} is relegated to Appendix A. For any given set of parameters, Lemma \ref{lm:size} reveals the fraction of individuals and cliques that are likely to
receive an information that is started from an arbitrary individual. Namely, an information
started from an arbitrary individual gives rise to an information epidemic with probability
$S_n$ (attributed to the possibility that the arbitrary node belongs to the giant component $G_H$),
and reaches a fraction $S_n$ of nodes in the network. Similar conclusions can be drawn in terms of $S_c$
for the fraction of cliques  that receive  the information.

Note that the condition (\ref{eq:threshold}) in Lemma \ref{lm:threshold}  depends on the
first/second moments of $d_w$ and $d_f$, which boils down to the linear combinations of the first/second moments of $k^w$ and $k^f$
in the following manner:
\begin{equation}
{\rm{E}}[{d_w}]  = \sum\limits_{n = 1}^D {\mu _n^wn{\rm{E}}[k_{}^w]}~~~~ {\rm{E}}[{d_f}]=\sum\limits_{m = 1}^D {\mu _m^fm{\rm{E}}[k_{}^f]},             \label{eq:E_d_w}
\end{equation}
\begin{equation}
 {\rm{E}}[{d_w}{d_f}]=  \sum\limits_{n = 1}^D {\sum\limits_{m = 1}^n {{\mu _{nm}}nm{\rm{E}}[k_{}^w]{\rm{E}}[k_{}^f]} },   \label{eq:E_dwdf}
\end{equation}
\begin{equation}
 {\rm{E}}[{({d_w})^2}] = \sum\limits_{n = 1}^D {\mu _n^w\left( {n{\rm{E}}[{{(k_{}^w)}^2}] + ({n^2} - n){{\left( {{\rm{E}}[k_{}^w]} \right)}^2}} \right)},    \label{eq:E_d_w2}
\end{equation}
\begin{equation}
 {\rm{E}}[{({d_f})^2}]= \sum\limits_{m = 1}^D {\mu _m^f\left( {m{\rm{E}}[{{(k_{}^f)}^2}] + ({m^2} - m){{\left( {{\rm{E}}[k_{}^f]} \right)}^2}} \right)}.     \label{eq:E_d_f2}
\end{equation}

According to Lemma  \ref{lm:size}, $S_c$ and $S_n$ are determined by ${\rm{E}}[h_1^{K_n^w}]$,
${\rm{E}}[K_n^wh_1^{K_n^w - 1}]$, ${\rm{E}}[h_2^{K_m^f}]$ and ${\rm{E}}[K_m^fh_2^{K_m^f - 1}]$
in (\ref{eq:h1})-(\ref{eq:node}), i.e., the integrals with respect to the distributions of $K_n^w$
and $K_m^f$  for different $n$ and $m$.
The calculations can be simplified by utilizing the following transformations:
 \begin{equation} \label{eq:E1}
 {\rm{E}}[h_1^{K_n^w}]  =  ( {{\rm{E}}[h_1^{{k^w}}]}  )^n ~~~ {\rm{E}}[h_2^{K_m^f}]  =  ( {{\rm{E}}[h_2^{{k^f}}]}  )^m,
  \end{equation}
 \begin{equation}  \label{eq:E2}
{\rm{E}}[K_n^wh_1^{K_n^w - 1}] = n{\left( {{\rm{E}}[h_1^{{k^w}}]} \right)^{n - 1}}{\rm{E}}[{k^w}h_1^{{k^w} - 1}],
 \end{equation}
 \begin{equation}  \label{eq:E3}
 {\rm{E}}[K_m^fh_2^{K_m^f - 1}] = m{\left( {{\rm{E}}[h_2^{{k^f}}]} \right)^{m - 1}}{\rm{E}}[{k^f}h_2^{{k^f} - 1}].
\end{equation}
With the help of (\ref{eq:E1})-(\ref{eq:E3}), we only need to calculate the integrals with respect to the distributions of  $k^w$ and $k^f$.
In this way, we can find $h_1$ and $h_2$ by numerically
solving the recursive equations (\ref{eq:h1})-(\ref{eq:h2})  and compute  $S_c$ and $S_n$ from (\ref{eq:gc})-(\ref{eq:node}), respectively.
The detailed derivations of (\ref{eq:E_d_w})-(\ref{eq:E3}) are omitted (see details in Appendix B).

\subsection{General Case: $0 \le T_w \le 1$ and $0 \le T_f \le 1$} \label{sec:sir}
We next generalize Lemma \ref{lm:threshold} and Lemma \ref{lm:size} to  the case
$0 \le T_w \le 1$ and $0 \le T_f \le 1$.   To this end, we  maintain the \emph{occupied} links in
the equivalent graph  $\mathbb{E}$
    by deleting each type-$1$ and type-$2$ edge with
probability $1-T_w$ and $1-T_f$, respectively. Let ${\tilde k^w}$ and ${\tilde k^f}$ be the
occupied link degrees (instead of $k^w$ and $k^f$) with the distributions $\{ \tilde p_k^w,
k=0,1,...\}$ and $\{\tilde p_k^f, k=0,1,...\}$. According to \cite{newman2002spread}, the
generating functions corresponding to ${\tilde k^w}$ and ${\tilde k^f}$ can be given by
\begin{equation} \label{eq:gen_sir}
 \tilde g(x) = g\left( {1 + {T_w}(x - 1)} \right)~~ \tilde q(x) = q\left( {1 + {T_f}(x - 1)} \right).
\end{equation}

From   (\ref{eq:E_d_w})-(\ref{eq:E3}),  we observe that the critical threshold and the giant component size are
determined by the distributions of  $k^w$ and $k^f$.
Therefore, Lemma \ref{lm:threshold} and Lemma \ref{lm:size} still hold if we replace the terms
associated with $k^w$ and $k^f$ in (\ref{eq:E_d_w})-(\ref{eq:E3}) by those associated with
${\tilde k^w}$ and ${\tilde k^f}$, respectively. To this end,
by using the generating functions (\ref{eq:gen_sir}), we find
\begin{eqnarray}  \nonumber
{\rm{E}}[{{\tilde k}^w}] &=& {T_w}{\rm{E}}[{k^w}],\\
{\rm{E}}[{({{\tilde k}^w})^2}] &=& T_w^2\left(
{{\rm{E}}[{{({k^w})}^2}] - {\rm{E}}[{k^w}]} \right) + {T_w}{\rm{E}}[{k^w}]. \nonumber
\end{eqnarray}
In the same manner, we can compute ${\rm{E}}[{{\tilde k}^f}]$ and ${\rm{E}}[{({{\tilde
k}^f})^2}]$. The critical threshold (in the general case) can now be computed by replacing
${\rm{E}}[{{ k}^w}]$, ${\rm{E}}[{{ k}^f}]$, ${\rm{E}}[{({{ k}^w})^2}]$,
${\rm{E}}[{({{ k}^f})^2}]$ with
${\rm{E}}[{{\tilde k}^w}]$, ${\rm{E}}[{{\tilde k}^f}]$, ${\rm{E}}[{({{\tilde k}^w})^2}]$,
${\rm{E}}[{({{\tilde k}^f})^2}]$, respectively, in (\ref{eq:E_d_w})-(\ref{eq:E_d_f2}).

In order to compute the giant component size,
we only need to  replace  the corresponding terms in (\ref{eq:E1})-(\ref{eq:E3})
with ${\rm{E}}[h_1^{\tilde k_{}^w}]$,  ${\rm{E}}[h_2^{\tilde k_{}^f}]$,  ${\rm{E}}[{{\tilde k}^w}h_1^{{{\tilde k}^w} - 1}]$  and ${\rm{E}}[{{\tilde k}^f}h_2^{{{\tilde k}^f} - 1}]$.
By using (\ref{eq:gen_sir}), we have
\[{\rm{E}}[h_1^{\tilde k_{}^w}] = \tilde g({h_1}) = {\rm{E}}[{(1 + {T_w}({h_1} - 1))^{{k^w}}}],\]
\[{\rm{E}}[{{\tilde k}^w}h_1^{{{\tilde k}^w} - 1}] = {\left[ {\tilde g({h_1})} \right]^\prime } = {T_w}{\rm{E}}[{k_w}{(1 + {T_w}({h_1} - 1))^{{k^w} -
1}}].\] Similar relations can be obtained for ${\rm{E}}[h_1^{\tilde k_{}^f}]$ and ${\rm{E}}[{{\tilde
k}^f}h_1^{{{\tilde k}^f} - 1}]$. The size of the giant component (in the general case)
can now be computed by reporting the
updated  (\ref{eq:E1})-(\ref{eq:E3}) into (\ref{eq:h1})-(\ref{eq:node}).

\section{Numerical results and simulations} \label{sec:numerical}
In this section, we numerically study the diffusion of real-time information by utilizing the
analytical results derived in  Section \ref{sec:analysis}. In particular, we are interested in how
the clique structure can impact the scale of information epidemic.  To get a
more concrete sense, we compare four system scenarios, each with different clique size distribution
as illustrated in Table \ref{tb:scenario}.

For the sake of fair comparison,  the total number of nodes in $\mathbb{W}$ is fixed at $12000$ in
each scenario. From  Table \ref{tb:scenario}, we can see that the average clique size  increases
from scenario $1$ to scenario $4$, indicating that individuals are getting closer to each other.
We assume that the type-$1$ link degree for each node in $\mathbb{W}$
 follows a poisson distribution, i.e., $p_k^w = \frac{{{\lambda ^k}}}{{k!}} \cdot {e^{ - \lambda }}$, $k = 0,1,2,...$,
where $\lambda$ is the average type-$1$ link degree. Meanwhile,  the type-$2$ link degree for each online user in $\mathbb{F}$ follows a power-law distribution with exponential cutoff, i.e., $p_{0}^f=0$ and
\begin{equation}
p_k^f = \frac{1}{C}{k^{ - \gamma }}{e^{ - \frac{k}{\Gamma }}}, \;\; k=1,2,\ldots,
\end{equation}
with the normalization factor $C = \sum\limits_{k = 1}^\infty  {{k^{ - \gamma }}{e^{ -
\frac{k}{\Gamma }}}}$.
\begin{table}
\caption{ The clique size distribution in four scenarios}
 \begin{center}
{
\begin{tabular}{|c|c|c|c|c|}
\hline      scenario  & size-$1$ & size-$2$ & size-$3$  & average clique size \\
\hline            $1$ & $100\%$         & $0$             & $0$             & $1$               \\
 \hline           $2$ & $66.7\%$         & $33.3\%$             & $0$             & $1.333$               \\
 \hline           $3$ & $33.3\%$         & $66.7\%$             & $0$             & $1.666$               \\
 \hline           $4$ & $33.3\%$         & $33.3\%$             & $33.3\%$             & $2$               \\
 \hline
\end{tabular}
} \vspace{-0.5cm}
\end{center}
\label{tb:scenario}
\end{table}
\begin{figure}[!th]
 \begin{center}
\includegraphics[width=0.3\textwidth]{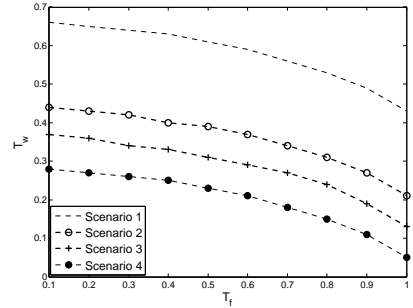}
 \end{center}
\caption{The minimum $T_w$ required for the existence of a giant component in  $\mathbb{E}$ versus
$T_f$ in four scenarios. We let $\lambda=1.5$, $\alpha=0.1$, $\gamma=3$ and $\Gamma=10$. Each curve
corresponds to the boundary of the phase transition in one scenario. A giant component is very
likely to emerge above the boundary.} \label{fig:critical}
\end{figure}
\begin{figure}[!t]
 \begin{center}
\includegraphics[width=0.27\textwidth]{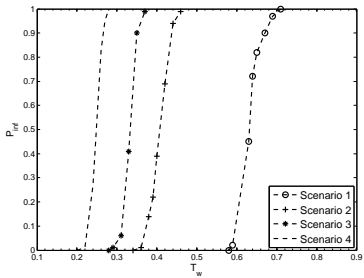}
 \end{center}
\caption{ The empirical probability $p_{inf}$ for the existence of giant component
is plotted.  The simulation results are obtained with $N=12000$ by averaging $200$ experiments.
We let  $T_f=0.4$ and other parameters follow the same setup as in Figure~\ref{fig:critical}.
 From scenario $1$ to scenario $4$, the values of $p_{inf}$ exhibit a sharp increase at
$T_w=0.64$, $T_w=0.4$, $T_w=0.35$ and $T_w=0.26$, respectively. Such sharp increase of $p_{inf}$ corresponds to the phase transition.   These values are  in good
agreement with the minimum required $T_w$ from the corresponding curves in
Figure~\ref{fig:critical}. We assume that a giant component exists if more than $5\%$ of the
cliques are connected.} \label{fig:pinf}
\end{figure}

We first compare these scenarios in term of the required conditions for the existence of a giant
component; in other words, in terms of the minimum conditions
for  an information epidemic to take place. We let $\lambda=1.5$,
$\alpha=0.1$, $\gamma=3$ and $\Gamma=10$. By computing the system's critical threshold, we depict in
Figure~\ref{fig:critical} the minimum required value of $T_w$ to have a giant component in $\mathbb{E}$
versus $T_f$. Each scenario corresponds to a  curve in the figure that stands for the boundary of
a phase transition; above the boundary a giant component is very likely to emerge. By the definition of
transmissibility, a lower required $T_w$ indicates that the system is more likely to give rise to a
giant component (and thus, to an information epidemic).
From scenario $1$ to scenario $4$, this figure clearly shows that
larger clique sizes lead to smaller values for the minimum $T_w$
required for an information epidemic, meaning that information
epidemics are more likely to take place for larger clique sizes.
The analytical results of Figure~\ref{fig:critical} are also
verified by simulations. For a fixed $T_f$, the probability of the
existence of giant component $p_{inf}$ is expected to have a sharp
increase as $T_w$ approaches to the corresponding minimum required
value in Figure~\ref{fig:critical}. Indeed, we observe in
Figure~\ref{fig:pinf} that when $T_f=0.4$, for each scenario, such
sharp transition occur at  $T_w$ close to the corresponding minimum
required value obtained in Figure~\ref{fig:critical}. Such sharp
increase of $p_{inf}$ corresponds to the phase transition, i.e.,  the giant
component could exist with high probability above the critical
threshold.  Therefore, the minimum required $T_w$ values obtained
via simulations are in good agreement with our analysis.

We next compare these scenarios in terms of the fractional sizes of influenced cliques and influenced
individuals.  For each scenario, we plot the fractional size of the giant component in $\mathbb{E}$
versus $T_f$ in Figure~\ref{fig:cliquesize}, which indicates the fraction of cliques that will
receive the information. We set $T_w=0.3$, $\lambda=2$, $\alpha=0.3$, $\gamma=3$ and $\Gamma=10$. In
this Figure, the curves stand for analytical results obtained by (\ref{eq:gc}), while the marked
points stand for the simulation results obtained by averaging $200$
experiments for each set of parameter. It is easy to check that the analytical results are in good agreement with the simulations. Obviously,  the fractional size of influenced cliques in scenario $4$ (with average clique size $2$) is
much larger than that in  scenario $1$ (with average clique size $1$), which  indicates that large cliques in the social network could greatly facilitate the message propagation.

We finally compare the fractional size of influenced individuals in Figure \ref{fig:nodesize}. In
this figure, the curves stand for the fractional size of influenced nodes obtained via
(\ref{eq:node}), whereas the marked points stand for the simulation results. Similar to
Figure~\ref{fig:cliquesize},  the information is much easier to propagate in a social network when larger size  cliques exist. For instance, when $T_f=1$, the fractional size of individuals
that receive the message grows sharply from $14\%$ (scenario $1$ with average clique size $1$) to
$80\%$ (scenario $4$ with average clique size $2$). In conclusion, the above results agree with a
natural conjecture that the messages are more influential (i.e., more likely to reach a large
portion of the population) when people are close to each other.
\begin{figure}[!t]
 \begin{center}
\includegraphics[width=0.3\textwidth]{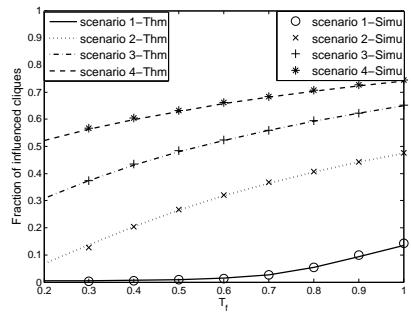}
 \end{center}
\caption{Fraction size of influenced cliques in  $\mathbb{W}$ versus $T_f$. We set $T_w=0.3$, $\lambda=2$, $\alpha=0.3$, $\gamma=3$ and $\Gamma=10$.
The curves stand for analytical results obtained by  (\ref{eq:gc}), while the marked points stand
for the simulation results with $N=12000$ by averaging $200$ experiments for each set of parameter. The analytical
results are in good agreement with the simulations.
 } \label{fig:cliquesize}
\end{figure}
\begin{figure}[!t]
 \begin{center}
\includegraphics[width=0.3\textwidth]{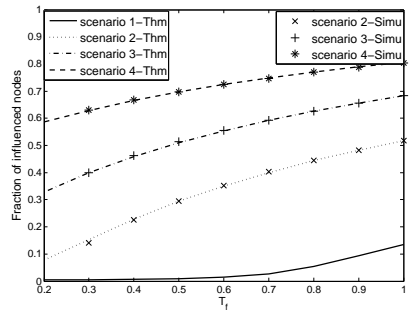}
 \end{center}
\caption{Fraction size of influenced nodes in  $\mathbb{W}$
versus $T_f$. All the parameters follow the same setup as in Figure~\ref{fig:cliquesize}.
 The curves stand for analytical results  obtained by (\ref{eq:node}) and the marked
points stand for the simulation results with $N=12000$.  The analytical
results are in good agreement with the simulations. For comparison, we also plot the fraction size of influenced cliques in scenario $1$ where the each clique has only one node.} \label{fig:nodesize}

\end{figure}

\section{conclusion}
In this study, we explore the diffusion of real-time information in social networks. We develop an overlaying  social-physical
network that consists of an online social network  and a physical information network with clique
structure. We theoretically quantify the condition and the size of information epidemics.
  To the best of our
knowledge, this paper is the first work on the diffusion of real-time information with
consideration on the clique structure in social networks. We believe that our findings  will
  offer initial steps towards understanding  the  diffusion behaviors of real-time information.

\section{Appendix}
\subsection{Proofs of Lemma \ref{lm:threshold} and Lemma \ref{lm:size}}
In \cite{Soderberg2,Soderberg4} S\"{o}derberg studied the phase transition behaviors of
inhomogeneous random graphs where nodes are connected  by different types of edges. Such graphs are
also called ``colored degree-driven random graphs'' in the sense that different types of edges
correspond to different colors. In a graph with $r$-types of edges, the edge degree of  an arbitrary node can be
represented by a $r$-dimension vector $\boldsymbol{d}=[d^{1} \:\: \cdots \:\: d^{r}]$, where
$d^{j}$ stands for the number of  type-$j$  edges incident on that node. In our study, the equivalent
graph $\mathbb{E}$ has two types of edges and the  degree distribution of an arbitrary node is
denoted  by $p(i,j) = {\rm{P[}}{d_w} = i,{d_f} = j{\rm{]}}$. Also, the generating function of
degree distribution $\{ p(i,j)\}$ can be defined by $H({x_1},{x_2}) = \sum\nolimits_i^{}
{\sum\nolimits_j^{} {p(i,j)x_1^ix_2^j} }$. Clearly, the multivariable combinatorial moments can be
achieved by partial differentiation at $x_1=1$ and $x_2=1$, i.e.,
 \begin{eqnarray}            \nonumber
 {\rm{E[}}{d_w}{\rm{]}} &=&  {\partial _1}H({x_1},{x_2}){|_{{x_1} = {x_2} = 1}},\\ \nonumber
 {\rm{E[}}{d_f}{\rm{]}} &=& {\partial _2}H({x_1},{x_2}){|_{{x_1} = {x_2} = 1}}, \\  \nonumber
 {\rm{E[}}{d_w}{d_f}{\rm{]}}&=&{\partial _1}{\partial _2}H({x_1},{x_2}){|_{{x_1} = {x_2} = 1}},\\  \nonumber
 {\rm{E[(}}{d_w}{{\rm{)}}^2}{\rm{]}}&=&\partial _1^2H({x_1},{x_2}){|_{{x_1} = {x_2} = 1}},\\  \nonumber
  {\rm{E[(}}{d_f}{{\rm{)}}^2}{\rm{]}}&=&   \partial _2^2H({x_1},{x_2}){|_{{x_1} = {x_2} = 1}}.
 \end{eqnarray}
Let $\{ {a_k}\}$ denote  size distribution of the largest connected component that can be
reached from an arbitrary node in   $\mathbb{E}$, whose generating function is defined by
 $g(z) = \sum\nolimits_k {{a_k}{z^k}}$.
Furthermore, we define a two-dimension vector $\mathbf{h}(z) = [{h_1}(z),{h_2}(z)]$, where
${h_i}(z)$ stands for the generating function of   size distribution of the component connected
by type-$i$ edges. According to the existing results in \cite{osmanmessage,Soderberg2,Soderberg4},
we have that
\begin{equation} \label{eq:gc_size}
g(z) = z\sum\limits_{i = 0}^\infty  {\sum\limits_{j = 0}^\infty  {p(i,j){h_1}{{(z)}^i}{h_2}{{(z)}^j} = zH(\mathbf{h}(z))} },
\end{equation}
where $\mathbf{h}(z)$ satisfies the  following recursive equations
 \begin{eqnarray}
 {h_1}(z) &=& \frac{z}{{{\rm{E[}}{d_w}{\rm{]}}}}{\partial _1}H(h(z)) \label{eq:req1},\\
 {h_2}(z) &=& \frac{z}{{{\rm{E[}}{d_f}{\rm{]}}}}{\partial _2}H(h(z))\label{eq:req2}.
 \end {eqnarray}
The emergence of giant component in $\mathbb{E}$ can be checked  by examining the stability  of
the recursive equations (\ref{eq:req1})-(\ref{eq:req2}) at the point ${h_1} = {h_1}(1) = 1$ and
${h_2} = {h_2}(1) = 1$. Along the same line as in \cite{Soderberg2,Soderberg4},
 we define a $2 \times 2$ Jacobian matrix $\mathbf{J}$, i.e.,
\[
\mathbf{J}= \left[ {\begin{array}{*{20}{c}}
   {{a_{11}}} & {{a_{12}}}  \\
   {{a_{21}}} & {{a_{22}}}  \\
\end{array}} \right],
\] where
\begin{eqnarray} \nonumber
{a_{11}} &=& \frac{1}{{{\rm{E[}}{d_w}{\rm{]}}}}\partial _1^2H(h(z)){|_{{h_1} = {h_2} = 1}} =
{{{\rm{E[(}}{d_w}{{\rm{)}}^2}{\rm{ - }}{d_w}{\rm{]}}} \mathord{\left/
 {\vphantom {{{\rm{E[(}}{d_w}{{\rm{)}}^2}{\rm{ - }}{d_w}{\rm{]}}} {{\rm{E[}}{d_w}{\rm{]}}}}} \right.
 \kern-\nulldelimiterspace} {{\rm{E[}}{d_w}{\rm{]}}}},\\ \nonumber
{a_{12}} &=& \frac{1}{{{\rm{E[}}{d_w}{\rm{]}}}}\partial _1^{}\partial _2^{}H(h(z)){|_{{h_1} = {h_2}
= 1}} = {{{\rm{E[}}{d_w}{d_f}{\rm{]}}} \mathord{\left/
 {\vphantom {{{\rm{E[}}{d_w}{d_f}{\rm{]}}} {{\rm{E[}}{d_w}{\rm{]}}}}} \right.
 \kern-\nulldelimiterspace} {{\rm{E[}}{d_w}{\rm{]}}}},\\ \nonumber
{a_{21}} &=& \frac{1}{{{\rm{E[}}{d_f}{\rm{]}}}}\partial _1^{}\partial _2^{}H(h(z)){|_{{h_1} = {h_2}
= 1}} = {{{\rm{E[}}{d_w}{d_f}{\rm{]}}} \mathord{\left/
 {\vphantom {{{\rm{E[}}{d_w}{d_f}{\rm{]}}} {{\rm{E[}}{d_f}{\rm{]}}}}} \right.
 \kern-\nulldelimiterspace} {{\rm{E[}}{d_f}{\rm{]}}}},\\ \nonumber
{a_{22}} &=& \frac{1}{{{\rm{E[}}{d_f}{\rm{]}}}}\partial _2^2H(h(z)){|_{{h_1} = {h_2} = 1}} =
{{{\rm{E[(}}{d_f}{{\rm{)}}^2}{\rm{ - }}{d_f}{\rm{]}}} \mathord{\left/
 {\vphantom {{{\rm{E[(}}{d_f}{{\rm{)}}^2}{\rm{ - }}{d_f}{\rm{]}}} {{\rm{E[}}{d_f}{\rm{]}}}}} \right.
 \kern-\nulldelimiterspace} {{\rm{E[}}{d_f}{\rm{]}}}}.
\end{eqnarray}
The spectral radius of $\mathbf{J}$ is given by \[\sigma=\frac{1}{2}\left( {{a_{11}} + {a_{22}} +
\sqrt {{{({a_{11}} - {a_{22}})}^2} + 4{a_{12}}{a_{21}}} } \right).
 \]
 By \cite{osmanmessage,Soderberg1,Soderberg4}, if $\sigma>1$,
 with high probability there exist a giant component in the graph $\mathbb{E}$;otherwise, a giant component is very
less likely to exist in $\mathbb{E}$. Therefore, the condition (\ref{eq:threshold}) in Lemma \ref{lm:threshold}
is achieved. Furthermore, the fraction size $S_c$ equals $1-g(1)$
\cite{osmanmessage}. By (\ref{eq:gc_size}), we have that
\begin{eqnarray}   \nonumber
{S_c} &=& 1 - g(1) = \sum\limits_{i = 0}^\infty  {\sum\limits_{j = 0}^\infty  {p(i,j)\left( {1 - h_1^ih_2^j} \right)} } \\
 &=& \sum\limits_{n = 1}^D {\sum\limits_{m = 0}^n {{\mu_{nm}}  {\sum\limits_{i = 0}^\infty  {\sum\limits_{j = 0}^\infty  {P_{ni}^wP_{mj}^f\left( {1 - h_1^ih_2^j} \right)} } }  } } \label{eq:Sc_Sn} \\  \nonumber
 &=& \sum\limits_{n = 1}^D {\sum\limits_{m = 0}^n {{\mu_{nm}}\left( {1 - \sum\limits_{i = 0}^\infty  {\sum\limits_{j = 0}^\infty  {P_{ni}^wP_{mj}^fh_1^ih_2^j} } } \right)} } \\   \nonumber
 &=& \sum\limits_{n = 1}^D {\sum\limits_{m = 0}^n {{\mu_{nm}}\left( {1 - {\rm{E}}[h_1^{K_n^w}]{\rm{E}}[h_2^{K_m^f}]} \right)} }.
\end{eqnarray}
In view of (\ref{eq:req1}) and (\ref{eq:req2}), we have that
\begin{eqnarray}  \nonumber
 {h_1}  &=& \frac{1}{{  {{\rm{E}}[{d_w}]}   }}\sum\limits_{i = 0}^\infty  {\sum\limits_{j = 0}^\infty{p(i,j)i h_1^{i - 1} h_2^j} }\\  \nonumber
 &=& \frac{1}{{ {{\rm{E}}[{d_w}]} }}\sum\limits_{n = 1}^D {\sum\limits_{m = 0}^n {{\mu _{nm}}{\rm{E}}[K_n^wh_1^{K_n^w - 1}]{\rm{E}}[h_2^{K_m^f}]} },
\end{eqnarray}
 \begin{eqnarray}  \nonumber
 {h_2} &=& \frac{1}{{ {{\rm{E}}[{d_f}] }   }}\sum\limits_{i = 0}^\infty  {\sum\limits_{j = 0}^\infty  {p(i,j)jh_1^ih_2^{j - 1}} }\\  \nonumber
 &=& \frac{1}{{{\rm{E}}[{d_f}] }}\sum\limits_{n = 1}^D {\sum\limits_{m = 0}^n {{\mu _{nm}}{\rm{E}}[h_1^{K_n^w}]{\rm{E}}[K_m^fh_2^{K_m^f - 1}]} }.
\end{eqnarray}
Furthermore, (\ref{eq:Sc_Sn}) can be rewritten in the following form:
\[
S_c = \sum\limits_{i = 0}^\infty  {\sum\limits_{j = 0}^\infty   \sum_{n=1}^{D}\sum_{m=0}^{n} \mu_{nm} P_{ni}^w P_{mj}^f { (1-h_1^ih_2^j)} }.
\]
Clearly, the term in parentheses gives the probability that a node with colored degree $[{d_w} =
i,{d_f} = j]$ belongs to the giant component. In other words, the term in parentheses is the
expected number of cliques added to the giant cluster by a degree $[{d_w} = i,{d_f} = j]$ clique.
 Hence, summing over all such $i,j$'s we get an expression for the {\em expected}
size of the giant cluster (in terms of number of cliques).

In order to compute the expected giant component size in terms of the number of nodes, namely to
compute $S_n$, we can modify the above expression such that the term  $n(1-h_1^i h_2^j)$  gives the
expected number of nodes to be included in the giant cluster by a degree $[{d_w} = i,{d_f} = j]$
clique. In other words, with probability $(1-h_1^i h_2^j)$ the clique under consideration will belong to the
giant component $G_H$ and will bring $n$ nodes to the actual giant size $S_n$. This yields
\begin{eqnarray}  \nonumber
{\bar{S}_n} &=& \sum\limits_{i = 0}^\infty  {\sum\limits_{j = 0}^\infty  {\sum\limits_{n = 1}^D {\sum\limits_{m = 0}^n { {\mu _{nm}}} } P_{ni}^wP_{mj}^f n\left( {1 - h_1^ih_2^j} \right)} } \\ \nonumber
 &=& \sum\limits_{n = 1}^D {\sum\limits_{m = 0}^n {n{\mu _{nm}}} } \sum\limits_{i = 0}^\infty  {\sum\limits_{j = 0}^\infty  {P_{ni}^wP_{mj}^f\left( {1 - h_1^ih_2^j} \right)} } \\ \nonumber
 &=& \sum\limits_{n = 1}^D {\sum\limits_{m = 0}^n {n{\mu _{nm}}\left( {1 - \sum\limits_{i = 0}^\infty  {\sum\limits_{j = 0}^\infty  {P_{ni}^wP_{mj}^fh_1^ih_2^j} } } \right)} } \\  \nonumber
 &=& \sum\limits_{n = 1}^D {\sum\limits_{m = 0}^n {n{\mu _{nm}}\left( {1 - {\rm{E}}[h_1^{K_n^w}]{\rm{E}}[h_2^{K_m^f}]} \right)}
 }.
\end{eqnarray}
We next have
\begin{equation}
{S_n} = \frac{1}{C}{\bar{S}_n},~~~C = \sum\limits_{n = 1}^D {n{\mu _n}},  \nonumber
\end{equation}
where the normalized term $C$ makes $S_n=1$ at $h_1=h_2=0$. Therefore, the conclusions  (\ref{eq:gc}) and
(\ref{eq:node}) in Lemma \ref{lm:size} have been obtained.

\subsection{Detailed Derivations for Equations (\ref{eq:E_d_w})-(\ref{eq:E3})}
As defined in Section \ref{sec:equal_graph}, $K_n^w$  is the sum of  $n$ independent copies of
$k^w$ and $K_m^f$ is the sum of  $m$ independent copies of $k^f$. It follows that
\begin{equation} \label{eq:mean}
 {\rm{E}}[K_n^w] = n{\rm{E}}[k_{}^w] ~~~~ {\rm{E}}[K_m^f] = m{\rm{E}}[k_{}^f],  \nonumber
\end{equation}
\begin{eqnarray}\nonumber
 {\rm{E}}[{(K_n^w)^2}] &=& {\mathop{\rm var}} [K_n^w] + {\left( {{\rm{E}}[K_n^w]} \right)^2}  \\  \nonumber
  &=& n{\rm{E}}[{(k_{}^w)^2}] + ({n^2} - n){\left( {{\rm{E}}[k_{}^w]} \right)^2} \label{eq:var1}, \\
 {\rm{E}}[{(K_m^f)^2}] &=& {\mathop{\rm var}} [K_m^f] + {\left( {{\rm{E}}[K_m^f]} \right)^2}  \nonumber \\
  &=& m{\rm{E}}[{(k_{}^f)^2}] + ({m^2} - m){\left( {{\rm{E}}[k_{}^f]} \right)^2}.  \label{eq:var2} \nonumber
\end{eqnarray}
In view of this, we can rewrite the first/second moments of $d_w$ and $d_f$ as follows:
\begin{eqnarray}\nonumber
 {\rm{E}}[{d_w}] &=& \sum\limits_{i = 0}^\infty  {\sum\limits_{j = 0}^\infty  {p(i,j)i} } \\
 &=& \sum\limits_{n = 1}^D {\sum\limits_{m = 0}^n {{\mu _{nm}}\left( {\sum\limits_{i = 0}^\infty  {\sum\limits_{j = 0}^\infty  {iP_{ni}^wP_{mj}^f} } } \right)} }  \nonumber  \\  \nonumber
  &=& \sum\limits_{n = 1}^D {\sum\limits_{m = 0}^n {{\mu _{nm}}{\rm{E}}[} K_n^w} ] = \sum\limits_{n = 1}^D {\mu _n^wn{\rm{E}}[k_{}^w]},
\end{eqnarray}
\begin{eqnarray}\nonumber
 {\rm{E}}[{d_f}] &=& \sum\limits_{i = 0}^\infty  {\sum\limits_{j = 0}^\infty  {p(i,j)j} } \\
  &=& \sum\limits_{n = 1}^D {\sum\limits_{m = 0}^n {{\mu _{nm}}\left( {\sum\limits_{i = 0}^\infty  {\sum\limits_{j = 0}^\infty  {jP_{ni}^wP_{mj}^f} } } \right)} }  \nonumber  \\ \nonumber
  &=& \sum\limits_{n = 1}^D {\sum\limits_{m = 0}^n {{\mu _{nm}}{\rm{E}}[} K_m^f} ] = \sum\limits_{m = 1}^D {\mu _m^fm{\rm{E}}[k_{}^f]},
\end{eqnarray}
\begin{eqnarray}\nonumber
 {\rm{E}}[{d_w}{d_f}] &=& \sum\limits_{i = 0}^\infty {\sum\limits_{j = 0}^\infty {p(i,j)i} } j \\ \nonumber
  &=&  \sum\limits_{n = 1}^D {\sum\limits_{m = 1}^n {{\mu _{nm}}nm{\rm{E}}[k_{}^w]{\rm{E}}[k_{}^f]} },  \label{eq:E_dwdf}
\end{eqnarray}
\begin{eqnarray}\nonumber
 {\rm{E}}[{({d_w})^2}] = \sum\limits_{i = 0}^\infty  {\sum\limits_{j = 0}^\infty  {p(i,j){i^2} = \sum\limits_{n = 1}^D {\mu _n^wE[{{(K_n^w)}^2}]} } }  \\ \nonumber
  = \sum\limits_{n = 1}^D {\mu _n^w\left( {n{\rm{E}}[{{(k_{}^w)}^2}] + ({n^2} - n){{\left( {{\rm{E}}[k_{}^w]} \right)}^2}} \right)},
\end{eqnarray}
\begin{eqnarray}\nonumber
 {\rm{E}}[{({d_f})^2}] = \sum\limits_{i = 0}^\infty  {\sum\limits_{j = 0}^\infty  {p(i,j){j^2} = \sum\limits_{m = 1}^D {\mu _m^fE[{{(K_m^f)}^2}]} } }  \\ \nonumber
  = \sum\limits_{m = 1}^D {\mu _m^f\left( {m{\rm{E}}[{{(k_{}^f)}^2}] + ({m^2} - m){{\left( {{\rm{E}}[k_{}^f]} \right)}^2}} \right)}.
\end{eqnarray}
We next characterize the generating functions of  $K_n^w$ and $K_m^f$. Specifically, the generating
functions of the type-$1$ and type-$2$ link degree distribution for a single node in $\mathbb{H}$
can be defined by $g(x) = \sum\nolimits_{k = 1}^\infty  {p_k^w{x^k}}$ and $q(x) = \sum\nolimits_{k
= 1}^\infty {p_k^f{x^k}}$. Since $K_n^w$ and $K_m^f$  are sums of i.i.d. random variables, their
 generating functions of turn out to be
\begin{equation} \label{eq:gf_n}
 {G_n}(x) = \sum\limits_{k = 0}^\infty  {P_{nk}^w{x^k}} = {\left[ {g(x)} \right]^n} ~~~~~ 1\le n \le D,
\end{equation}
\begin{equation}\label{eq:gf_m}
 {Q_m}(x) = \sum\limits_{k = 0}^\infty  {P_{mk}^f{x^k}} = \left\{ {\begin{array}{*{20}{c}}
   {{{\left[ {q(x)} \right]}^m}} ~~~  1 \le m \le D, \\
   0  ~~~~~~~~~~ m=0\\
\end{array}} \right.
\end{equation}
With (\ref{eq:gf_n}) and (\ref{eq:gf_m}), ${\rm{E}}[h_1^{K_n^w}]$, ${\rm{E}}[K_n^wh_1^{K_n^w -
1}]$, ${\rm{E}}[h_2^{K_m^f}]$ and ${\rm{E}}[K_m^fh_2^{K_m^f - 1}]$ can boil down to the expected
value with respect to the distribution of  $k^w$ and $k^f$ as follows.
\begin{equation}
\begin{array}{l}  \nonumber
 {\rm{E}}[h_1^{K_n^w}] = {G_n}({h_1}) = {\left( {g({h_1})} \right)^n} =  ( {{\rm{E}}[h_1^{{k^w}}]}  )^n  \\
 {\rm{E}}[h_2^{K_m^f}] = Q_m^{}({h_2}) = {\left( {q({h_2})} \right)^m} =  ( {{\rm{E}}[h_2^{{k^f}}]}  )^m
 \end{array}
  \end{equation}
 \begin{eqnarray} \nonumber
{\rm{E}}[K_n^wh_1^{K_n^w - 1}] &=& {G_n}({h_1})' = n{\left( {g({h_1})} \right)^{n - 1}}{\left( {g({h_1})} \right)^\prime } \\  \nonumber
  &=& n{\left( {{\rm{E}}[h_1^{{k^w}}]} \right)^{n - 1}}{\rm{E}}[{k^w}h_1^{{k^w} - 1}]
 \end{eqnarray}
 \begin{eqnarray}  \nonumber
 {\rm{E}}[K_m^fh_2^{K_m^f - 1}] &=& {Q_m}({h_2})' = m{\left( {q({h_2})} \right)^{m - 1}}{\left( {q({h_2})} \right)^\prime } \\  \nonumber
  &=& m{\left( {{\rm{E}}[h_2^{{k^f}}]} \right)^{m - 1}}{\rm{E}}[{k^f}h_2^{{k^f} - 1}].
\end{eqnarray}

{
\bibliographystyle{unsrt}
\bibliography{reference}
}

\end{document}